\documentclass[aps,preprint]{revtex4}%
\usepackage{amsfonts}
\usepackage{amsmath}
\usepackage{amssymb}
\usepackage{graphicx}%
\begin{document}
\title{Transfer of Gravitational Information through a Quantum Channel}
\author{Baocheng Zhang}
\email{zhangbc@wipm.ac.cn}
\author{Qing-yu Cai}
\author{Ming-sheng Zhan}
\affiliation{State Key Laboratory of Magnetic Resonances and Atomic and Molecular Physics,
Wuhan Institute of Physics and Mathematics, Chinese Academy of Sciences, Wuhan
430071, People's Republic of China}

\begin{abstract}
Gravitational information is incorporated into an atomic state by correlation
of the internal and external degrees of freedom of the atom, in the present
study of the atomic interferometer. Thus it is difficult to transfer
information by using a standard teleportation scheme. In this paper, we
propose a novel scheme for the transfer of gravitational information through a
quantum channel provided by the entangled atomic state. Significantly, the
existence of a quantum channel suppresses phase noise, improving the
sensitivity of the atomic interferometer. Thus our proposal provides novel
readout mechanism for the interferometer with an improved signal-to-noise ratio.



\end{abstract}
\maketitle




\section{Introduction}

One of the most important recent achievements regarding the incorporation of
gravity into quantum theory, is the measurement of gravitational effects by
using quantum states of microscopic particles \cite{cow75,rkb91,kc91}, since
gravitational forces are too weak to be readily observable on the microscopic
scale. In current experiments, especially for the atomic interferometer
\cite{rkb91,kc91}, the probe of gravitational effects is dependent on the
change of atomic internal state, which is measured using the interference
effects involving both the internal and external degrees of freedom. Moreover,
due to the technological developments of laser cooling and atom trapping, more
elaborate effects due to the coupling of gravity with the atomic state will be
detected by the atomic interferometer in the near future \cite{dgk07,dgk08}.
On the other hand, it is usually necessary to measure gravitational field in
places inaccessible to people. Then, when the atomic interferometer is sent to
the appointed location, we have to find a method to read the measured result.
Although there might be many other methods that solve this problem, in this
paper we address whether gravitational information written into the atomic
state can be transferred through a quantum channel. At the same time our study
also involves another problem of whether the information stored in the
entangled quantum state of a single atom by correlation of its internal and
external degrees of freedom (for the sake of convenience we call such a state
a \textquotedblleft specific state\textquotedblright\ in this paper) can be
transferred through the quantum channel, which has not been studied up to now
as far as we know.

The past few years have seen much active study of information transfer since
the emergence of quantum entanglement as a channel for the transfer of
information, and most of these studies focus on a method called quantum
teleportation \cite{bbw93,ncc00,ncc09}. Thus a natural and direct thought is
that whether we can use the process of quantum teleportation to implement the
transfer of classical gravitational information, since this information can be
stored in an atomic quantum state. Actually such a process is not feasible.
The reason for this is that the atomic quantum state is formed by correlation
of the internal and external degrees of freedom of a single atom that goes
through the atomic interferometer. Gravitational information will be stored in
the \textquotedblleft specific state\textquotedblright\ which makes the
standard teleportation process invalid since a joint measurement cannot be
made in this case.

In order to realise the transfer of classical gravitational information stored
in the \textquotedblleft specific state\textquotedblright, we propose a novel
method in which two inter-entangled atoms are used, but the entanglement
exists only between the internal atomic degrees of freedom. Thus the entangled
atomic state provides a quantum channel through which gravitational
information can be transferred by writing them into one part of the entangled
state. However, the question of whether the existence of this quantum channel
will change the sensitivity of the results for testing gravitational effects
must be investigated carefully. In this paper, we put forward a practical
scheme for the transfer of information, and analyze the influence of the
existence of the quantum channel on the results. Our analysis reveals that the
mechanism scales better with respect to phase noise than the previous one
without the existence of the quantum channel. Thus our proposal provides novel
readout mechanism for the interferometer, with a better signal-to-noise ratio.

The structure of this paper is as follows. We next describe how gravitational
information is stored in the \textquotedblleft specific
state\textquotedblright\ in the second section. Then the third section is
devoted to the investigation of the transfer of information about the
classical gravitational field through quantum channel. In the fourth section,
we compare the influence of noise on gravitational information for the two
situations, with and without the existence of quantum channel, and find that
the existence of the quantum channel improves the measurement sensitivity of
the atomic interferometer. Finally, we discuss and summarise our results in
the fifth section.

\section{Specific State}

In the above section, we have introduced the \textquotedblleft specific
state\textquotedblright\ which involves the correlation of the internal and
external degrees of freedom of a single atom. In this section we will present
the \textquotedblleft specific state\textquotedblright, without loss of
generality, within the structure of the atomic interferometer proposed firstly
by M. Kasevich and S. Chu \cite{kc91}. Due to the absence of a complete
quantum gravity theory, the interferometry process is generally described
semiclassically, which means the gravitational field is considered to be
classical while the motion of the matter is described according to quantum mechanics.

The interferometer considered here mainly consists of a
beamsplitter-mirror-beamsplitter ($\frac{\pi}{2}-\pi-\frac{\pi}{2}$) optical
pulse sequence and is the matter-wave analog of a Mach-Zender interferometer.
See Fig.1 for a description of the principle of the interferometer, which can
generally be stated as such: Firstly, an atom, prepared in the ground state
$\left\vert g\right\rangle $, interacts with the first $\frac{\pi}{2}$ pulse,
which will couple the two stable internal atomic states $\left\vert
g\right\rangle $ and $\left\vert e\right\rangle $. Then the atomic state
becomes a coherent superposition of the two states $\left\vert g\right\rangle
$ and $\left\vert e\right\rangle $, with the corresponding changes to the
external states, which leads to separation of the atoms on the basis of the
momentum of the photon used to drive the transition. After a period of free
flight of the atoms, the $\pi$ pulse forces the exchange of the internal
states associated with the atomic wavepackets, and the simultaneous exchange
of the external states through the photon recoil momentum. Finally, the second
$\frac{\pi}{2}$ pulse induces the interference between the atomic wavepackets
along different recoil paths, which leads to a change of the probability of
finding the atom in the one of the two states $\left\vert g\right\rangle $ and
$\left\vert e\right\rangle $, and gives the observed signal in the experiment.
To sum up, we equivalently express the interference process by solving the
Schr\"{o}dinger equation with the following Hamiltonian \cite{cl96,dgk08}:%

\begin{equation}
H=\frac{p^{2}}{2m}+%
{\displaystyle\sum\limits_{i}}
E_{i}\left\vert A_{i}\right\rangle \left\langle A_{i}\right\vert
-\mathbf{d}\cdot\mathbf{E+}V\left(  x\right)  \label{ham}%
\end{equation}
where the first term is the kinetic energy of the atoms (which will influence
the external degrees of freedom), the second term is the internal atomic
structure Hamiltonian, the third term is the atom-light interaction (with
$\mathbf{d}$ the electric dipole operator and $\mathbf{E}$ the electric
field), and the last term describes the interaction with the gravitational
field which can be expanded up to the second order as, $V\left(  r\right)
=V\left(  r_{0}\right)  -g_{a}\left(  r_{a}-r_{0}\right)  +\frac{1}{2}%
V_{ab}\left(  r_{a}-r_{0}\right)  \left(  r_{b}-r_{0}\right)  $, with the
gravitational acceleration $g_{a}=-\frac{1}{m}\partial_{a}V\left(
r_{0}\right)  $ and the second derivative $V_{ab}=\partial_{a}\partial
_{b}V\left(  r_{0}\right)  $ (which is related to the gravitational gradient
or the Riemannian space-time curvature). Then we calculate the evolution of
the state in the interaction picture, using a decomposition of the state
vector of the free evolution into plane waves,%
\begin{equation}
\left\vert \chi\right\rangle =%
{\displaystyle\int}
d\mathbf{p}%
{\displaystyle\sum\limits_{i}}
c_{i}\left(  \mathbf{p},t\right)  e^{-i\left(  E_{i}+\frac{p^{2}}{2m}\right)
\left(  t-t_{0}\right)  }\left\vert \mathbf{p}\right\rangle \left\vert
A_{i}\right\rangle
\end{equation}
where $\left\vert \mathbf{p}\right\rangle $ is the eigenvector of the atom's
external degrees of freedom with the momentum $\mathbf{p}$. Then according to
the principle of the atomic interferometer, after all interactions, we have
the state
\begin{equation}
\left\vert \chi\right\rangle =\frac{1}{2}\left[  \left(  1+e^{-i\Delta
\phi_{tot}}\right)  \left\vert g\right\rangle \left\vert p\right\rangle
+\left(  1-e^{-i\Delta\phi_{tot}}\right)  \left\vert e\right\rangle \left\vert
p+k\right\rangle \right]  \label{ss}%
\end{equation}
where $\Delta\phi_{tot}$ is the total phase difference
\cite{dgk08,kc92,sc94,pcc97,dgk07,ysd03} between two paths through the atomic
interferometer, which provides the basis of an experimental observation. The
state in Eq. (\ref{ss}) is the \textquotedblleft specific
state\textquotedblright\ that we are concerned with in this paper which can
evidently be seen the correlation between the internal and external degrees of
freedom. In addition, the state in Eq. (\ref{ss}) also includes information
about the gravitational field around the atomic interferometer. Assuming that
the duration of atom-laser interaction is short compared with the duration of
the free-flight time of the atom and there is little dispersion of the atomic
wavepacket in the whole interference process, the total phase difference is
expressed \cite{kc92,pcc97} as%
\begin{equation}
\Delta\phi_{tot}=kgT^{2} \label{phase}%
\end{equation}
where $k$ is the effective laser-field wavevector, $T$ is the interrogation
time between two subsequent laser pulses, and $g$ is the information from the
gravitational field, conditional on the experimental setup and precision.
Actually, the gravitational field around the Earth is complicated and its
complication has been partly observed in recent experiments. The present
precision for the value of the gravitational acceleration is $\Delta
g/g\simeq3\times10^{-9}$, the measurement of which involved gaining
information about the gravity gradient by the corresponding measurement of the
phase difference \cite{pcc99,zwz11},
\begin{equation}
\Delta\phi_{tot}=\left(  1+\frac{7}{12}\gamma T^{2}+\cdots\right)  kg_{0}%
T^{2},
\end{equation}
where $\gamma\simeq3\times10^{-7}gm^{-1}$ is the gradient of the Earth's
gravitational field. More effects due to general relativity, such as
non-linear three-graviton coupling, the gravity of kinetic energy, and so on,
have been analysed using the same kind of atomic interferometry and the
relevant experimental setup is presently going on \cite{dgk08,zxwz11}. Even
information about gravitational waves from distant astrophysical events
\cite{cs04,gjz11} can also be tested using such atomic interferometry, but
there is a need for higher precision and accuracy.

Finally, when the atom leaves the interferometer, the probability of finding
it still in the ground state is \cite{dgk08,cl96}%
\begin{equation}
P_{t}=\frac{1+\cos(\Delta\phi_{tot})}{2}\label{pr}%
\end{equation}
Thus, the gravitational information is written into the atomic population,
which gives the experimental signal. So the successful transfer of the
classical gravitational information is equivalent to finding a similar
population in a location far from the interferometer. On the other hand, it is
noted that the state in Eq. (\ref{ss}) shows the coupling of the internal and
external degrees of freedom. Thus the transfer of information is closely
related to the treatment of the \textquotedblleft specific
state\textquotedblright.

\section{Gravitational Information Transfer}

As stated in the last section, if we want to transfer classical gravitational
information, the \textquotedblleft specific state\textquotedblright\ has to be
treated carefully. Due to the coupling of the internal and external degrees of
freedom, in order to implement the transfer our strategy is such: At first we
will prepare an entangled atomic state between only the internal degrees of
freedom of two atoms, and then make one part of the entangled state go through
the atomic interferometer to carry the information about the gravitational
field, and make the other part maintain its initial situation. The interaction
of the atom with the gravitational field leads to the coupling between its
internal and external degrees of freedom, and at the same time changes the
internal entanglement of the atom with the other atom; in this way
gravitational information is transferred. When the part going through the
interferometer finishes its interaction, we perform a velocity selection on
these atoms. After all these operations, we measure the other part placed in
the original location in order to to obtain the gravitational information. For
clarity, see Fig.1. In what follows, we will implement this strategy.

\begin{figure}[p]
\centering
\includegraphics[width=1.0\textwidth,bb=0 0 600 600]{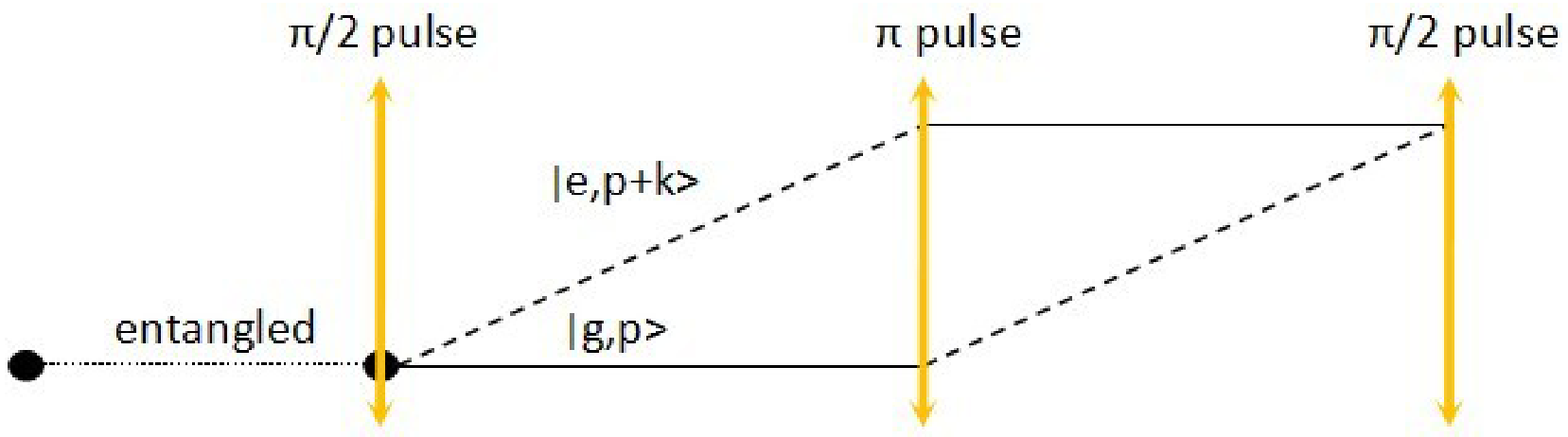}\caption{
Sketch of our scheme. The left sector is the quantum channel
provided by the most entangled atomic pairs. The right sector is the
schematic of the atomic
interferometer.}%
\label{Fig.1}%
\end{figure}

The initial entangled atomic state prepared is%
\begin{equation}
\left\vert \Psi\right\rangle =\frac{1}{\sqrt{2}}\left(  \left\vert
g\right\rangle _{1}\left\vert e\right\rangle _{2}+\left\vert e\right\rangle
_{1}\left\vert g\right\rangle _{2}\right)  \label{es}%
\end{equation}
where $\left\vert g\right\rangle $ represents the ground state, $\left\vert
e\right\rangle $ represents the excited state and the subscripts label the two
entangled atoms. Here, we introduce a method regarding the preparation of an
entangled state by the exchange of a single photon between the atoms in a high
Q cavity, based on that of Ref. \cite{cz94,hmh97}. In such a preparation the
atom-cavity system undergoes oscillations, with Rabi frequency $\Omega$,
between states $\left\vert g\right\rangle \left\vert 1\right\rangle
,\left\vert e\right\rangle \left\vert 0\right\rangle $ which means that the
atom in the internal state $\left\vert g\right\rangle $ or $\left\vert
e\right\rangle $ coexists with one photon or zero photon in the cavity. The
process can be described simply as follows: The first atom in a state
$\left\vert e\right\rangle $ enters and couples with the cavity with a
duration $\Omega t_{1}=\frac{\pi}{2}$, and then the second atom in a state
$\left\vert g\right\rangle $ enters the cavity. Before the second atom enters,
the state of the system consisting of two atoms and the cavity is%
\begin{equation}
\left\vert \Psi^{^{\prime}}\right\rangle =\frac{1}{\sqrt{2}}\left(  \left\vert
g\right\rangle _{1}\left\vert 1\right\rangle +\left\vert e\right\rangle
_{1}\left\vert 0\right\rangle \right)  \left\vert g\right\rangle _{2}%
\end{equation}
After a delay $T$, the second atom enters and couples with the cavity with a
duration $\Omega t_{2}=\pi$. If the first atom has left the cavity empty, the
second atom will stay in the initial state. If the first atom has emitted a
photon, the second atom will absorb it and change to the state $\left\vert
e\right\rangle $. As a result, the final state becomes%
\begin{equation}
\left\vert \Psi^{^{\prime}}\right\rangle =\frac{1}{\sqrt{2}}\left(  \left\vert
g\right\rangle _{1}\left\vert e\right\rangle _{2}+\left\vert e\right\rangle
_{1}\left\vert g\right\rangle _{2}\right)  \left\vert 0\right\rangle
=\left\vert \Psi\right\rangle \left\vert 0\right\rangle
\end{equation}
So the entangled atomic state appears in the presence of an empty cavity in
the same form as that of Eq. (\ref{es}). Notably, such an entangled atomic
state has been realised experimentally with good fidelity \cite{hmh97,rnh00}.
However, the entangled atomic state is fragile due to detrimental effects of
the environment, so the two entangled atoms cannot be separated too far.
Fortunately, a recent study regarding the dissipative preparation of
entanglement in optical cavities showed that the environment can be used as a
resource to help generate entanglement \cite{krs11}. In the study two
$\Lambda$-type three-level atoms were entangled in a detuned cavity with the
interaction initiated by a off-resonance optical laser and a microwave source.

With the entangled atomic state of Eq. (\ref{es}), our scheme can be
implemented. Firstly we make the second atom undergo the interference process
in the interferometer described in the last section, and according to the
Hamiltonian in Eq. (\ref{ham}), the change of its internal and external atomic
states is expressed as%
\begin{align*}
\left\vert g\right\rangle \left\vert p\right\rangle  &  \rightarrow
a_{1}\left\vert g\right\rangle \left\vert p\right\rangle +a_{2}\left\vert
e\right\rangle \left\vert p+k\right\rangle \\
\left\vert e\right\rangle \left\vert p\right\rangle  &  \rightarrow
b_{1}\left\vert g\right\rangle \left\vert p-k\right\rangle +b_{2}\left\vert
e\right\rangle \left\vert p\right\rangle
\end{align*}
with the parameters calculated by%
\begin{align*}
a_{1}  &  =-\frac{1}{2}e^{-i\phi_{2}+i\phi_{1}}\left(  1+e^{-i\Delta\phi
}\right) \\
a_{2}  &  =\frac{i}{2}e^{-i\phi_{2}+i\phi_{1}+i\phi_{3}}\left(  1-e^{-i\Delta
\phi}\right) \\
b_{1}  &  =\frac{i}{2}e^{i\phi_{2}-i\phi_{1}-i\phi_{3}}\left(  1-e^{i\Delta
\phi}\right) \\
b_{2}  &  =-\frac{1}{2}e^{i\phi_{2}-i\phi_{1}}\left(  1+e^{i\Delta\phi
}\right)
\end{align*}
where $\phi_{1}$, $\phi_{2}$ and $\phi_{3}$ are the initial phases of the
$\frac{\pi}{2}-\pi-\frac{\pi}{2}$ laser pulse sequence, and $\Delta\phi=$
$\left(  \phi_{1}\left(  t_{1}\right)  -\phi_{2}\left(  t_{2}\right)  \right)
_{up-path}-\left(  \phi_{3}\left(  t_{3}\right)  -\phi_{2}\left(
t_{2}\right)  \right)  _{down-path}$ where $t_{i}$ is the time of the
light-atom interaction. The phase difference existing in the atomic
interferometer is calculated according to the expression
\cite{dgk08,kc92,sc94,pcc97,dgk07} $\Delta\phi=\Delta\phi_{path}+\Delta
\phi_{light}+\Delta\phi_{seperation}$, where the phase difference $\Delta
\phi_{path}$ originates from the free-flight evolution of the atom between
light pulses, the phase difference $\Delta\phi_{light}$ comes from the
interaction of the atom with the laser field used to manipulate the
wave-function at each of the beamsplitters and mirrors in the interferometer,
and the phase difference $\Delta\phi_{seperation}$ arises due to the final
spatial separation of the interfering atomic wavepackets at the interferometer
output port. Thus information about the gravitational field involved in the
process is printed on the phase difference $\Delta\phi$ or the parameters
$a_{1},a_{2},b_{1},b_{2}$.

So if the second atom goes through the process in the atomic interferometer,
after the interaction, the whole state becomes%
\begin{align}
\left\vert \Psi_{t}\right\rangle  &  =\frac{1}{\sqrt{2}}\left(  \left\vert
g\right\rangle _{1}\left(  b_{1}\left\vert g\right\rangle _{2}\left\vert
p-k\right\rangle +b_{2}\left\vert e\right\rangle _{2}\left\vert p\right\rangle
\right)  +\left\vert e\right\rangle _{1}\left(  a_{1}\left\vert g\right\rangle
_{2}\left\vert p\right\rangle +a_{2}\left\vert e\right\rangle _{2}\left\vert
p+k\right\rangle \right)  \right)  \nonumber\\
&  =\frac{1}{\sqrt{2}}\left(  b_{1}\left\vert g\right\rangle _{1}\left\vert
g\right\rangle _{2}\left\vert p-k\right\rangle +(b_{2}\left\vert
g\right\rangle _{1}\left\vert e\right\rangle _{2}+a_{1}\left\vert
e\right\rangle _{1}\left\vert g\right\rangle _{2}\right)  \left\vert
p\right\rangle +a_{2}\left\vert e\right\rangle _{1}\left\vert e\right\rangle
_{2}\left\vert p+k\right\rangle )
\end{align}
When the atom leaves the interferometer, velocity selection has to be made
again using the same procedure as for the atom entering into the
interferometer \cite{dgk08,zxwz11}. If the external state is also projected
onto the eigenvector $\left\vert p\right\rangle $ with the same momentum $p$
as that of the entering atoms, the probability of measuring the state
$\left\vert g\right\rangle _{1}$ is%
\begin{equation}
P=\frac{1+\cos(\Delta\phi)}{4}\label{tp}%
\end{equation}
which is different from the result presented in Eq. (\ref{pr}). However, since
the gravitational information is stored in the total phase difference
$\Delta\phi$, the transfer is still complete although the observed fringe
amplitude changes. Thus we realise the transportation of classical
gravitational information through a quantum channel. In the following section,
we will discuss the difference between the two results of Eqs. (\ref{pr}) and
(\ref{tp}) and the influence of the existence of a quantum channel on the
measurement sensitivity.

\section{Influence of The Quantum Channel}

It is noted that gravitational information is being stored in the total phase
difference $\Delta\phi$ when the information is transferred, which shows that
gravitational information will not be lost in the process of transfer through
the quantum channel. But the existence of the quantum channel has an influence
on the measurement of gravitational information by the second atom going
through the atomic interferometer, \textit{i.e.} the observed fringe amplitude
in Eq. (\ref{tp}) becomes $\frac{1}{4}$ instead of the original amplitude
$\frac{1}{2}$\ in Eq. (\ref{pr}), given a fringe contrast of $100\%$. So it is
very interesting to see what role is played by the noise of the interferometer
in the process of information transfer. Or equivalently, whether the quantum
channel will limit the influence of noise on the measurement process, since
the information about noise is also carried together with gravitational
information in the transfer process. Alongside the analysis made in Ref.
\cite{ysd03}, we also carry out a simple calculation for the signal-to-noise
ratio with and without the existence of the quantum channel.

Here, all we consider is the measurement of the gravitational acceleration
which is expressed as
\begin{equation}
\Delta\phi=\Delta\phi_{tot}=kgT^{2}+O(T^{3})
\end{equation}
where $O(T^{3})$ contains the information about the higher-order correction of
gravitational acceleration. Although gravitational information is stored in
the total phase difference, the real measurement is of the atomic population,
so we have to consider firstly the influence of change of atomic population on
the results of the measurement result which is also called as shot noise.
Generally, we could express the atomic population as%
\begin{equation}
P_{t}=\frac{1+\cos(\Delta\phi_{tot})}{2}=\frac{N_{t}}{N}%
\end{equation}
without the existence of the quantum channel, and
\begin{equation}
P=\frac{1+\cos(\Delta\phi)}{4}=\frac{N_{1}}{N}%
\end{equation}
with the existence of the quantum channel, where $N_{t}$ and $N_{1}$ are the
number of ground-state atoms at the end of the interrogation process for the
two different situations, without and with the quantum channel, respectively.
Thus assumpting a binomial distribution for the random variables, for $N$
trials and with probabilities $P_{t}$ and $P$, we estimate the shot noise
contribution to each phase measurement by the formula%
\begin{equation}
P_{ts}=\frac{\partial\Delta\phi_{tot}}{\partial N_{t}}\sigma_{N_{t}}=\frac
{1}{\sqrt{N}}%
\end{equation}
without the existence of the quantum channel, and
\begin{equation}
P_{s}=\frac{\partial\Delta\phi}{\partial N_{1}}\sigma_{N_{1}}=\sqrt{\frac
{2}{N}}%
\end{equation}
with the existence of the quantum channel, where $\sigma_{N_{t}}$ and
$\sigma_{N_{1}}$ are ordinary variances defined as $\sigma_{X}=\sqrt
{\left\langle X^{2}\right\rangle -\left\langle X\right\rangle ^{2}}$. Thus we
find that the existence of the quantum channel will amplify the influence of
shot noise on the measurement by
\begin{equation}
\frac{P_{s}}{P_{ts}}=\sqrt{2}.\label{sn}%
\end{equation}

However, in the following we find that the phase noise is suppressed by the
existence of the quantum channel. Here, we analyze the phase noise using the
same model in which the noise is simply regarded as proportional to the
magnitude of the fluorescence signal \cite{ysd03,smk98}. According to the
model, we get%
\begin{equation}
\left\langle \Delta\phi_{tot}^{2}\right\rangle -\left\langle \Delta\phi
_{tot}\right\rangle ^{2}=c^{2}\frac{1+\cos(\left\langle \Delta\phi
_{tot}\right\rangle )}{2}%
\end{equation}
without the existence of the quantum channel, and
\begin{equation}
\left\langle \Delta\phi^{2}\right\rangle -\left\langle \Delta\phi\right\rangle
^{2}=c^{2}\frac{1+\cos(\left\langle \Delta\phi\right\rangle )}{4}%
\end{equation}
with the existence of the quantum channel, where $c$ is a constant. Then the
phase noise contribution to each phase measurement is%
\begin{equation}
P_{tp}=c\sin\frac{\left\langle \Delta\phi_{tot}\right\rangle }{2}%
\end{equation}
without the existence of the quantum channel, and
\begin{equation}
P_{p}\simeq\frac{1}{\sqrt{2}}c\sin\frac{\left\langle \Delta\phi\right\rangle
}{2}%
\end{equation}
with the existence of the quantum channel, and their ratio is
\begin{equation}
\frac{P_{p}}{P_{tp}}\simeq\frac{1}{\sqrt{2}}. \label{pn}%
\end{equation}
From the comparisons of the contributions to the noise given by Eqs.
(\ref{sn}) and (\ref{pn}), we find that shot noise is amplified due to the
loss of atoms, but phase noise is suppressed when gravitational information is
transferred through the quantum channel. It is noted that the influence on the
measurement of the total phase difference of phase noise is about a hundred
times larger than that of shot noise in the measurement of gravitational
acceleration presented in Ref. \cite{pcc99}, so the net result for the
measurement sensitivity is that gravitational information will be affected
less by noise when the quantum channel exists compared to without the
existence of the quantum channel. Thus we conclude that the gravitational
information transferred through the quantum channel will be purer than that
measured directly, since the noise will be suppressed overall due to the
existence of the quantum channel.

Then, there is still the problem of whether the quantum channel, as described
by Eq. (\ref{es}), is optimal for suppression of the phase noise in our scheme
of information transfer. For answering this problem, we take the general
entangled state%

\begin{equation}
\left\vert \Phi\right\rangle =a\left\vert g\right\rangle _{1}\left\vert
e\right\rangle _{2}+b\left\vert e\right\rangle _{1}\left\vert g\right\rangle
_{2} \label{exg}%
\end{equation}
where $\left\vert a\right\vert ^{2}+\left\vert b\right\vert ^{2}=1$. Then,
implementing our scheme in the same way as in the above section, we obtain the
final probability
\[
P_{\Phi}=\frac{1+\cos(\Delta\phi)}{2}\left\vert a\right\vert ^{2}.
\]
Thus the ratio of Eq. (\ref{pn}) changes into
\begin{equation}
\frac{P_{\Phi p}}{P_{tp}}\simeq\sqrt{\left\vert a\right\vert \left\vert
b\right\vert }. \label{png}%
\end{equation}
From the perspective of suppressing phase noise, the optimal selection is
$\left\vert a\right\vert =\left\vert b\right\vert =\frac{1}{\sqrt{2}}$ as
found by calculating the extremum of Eq. (\ref{png}) with the normalisation
relation between $a$ and $b$. However, we have to investigate whether this is
also optimal for the transfer of information. In order to estimate the
selection for optimal transfer of information, we have to maximise the
gravitational information obtained after transfer by using the entropy
$S=\sum_{i}P_{i}\log P_{i}$ with the probabilities $P_{1}=$ $\frac
{1+\cos(\Delta\phi)}{2}\left\vert a\right\vert ^{2}$ and $P_{2}=$
$\frac{1-\cos(\Delta\phi)}{2}\left\vert b\right\vert ^{2}$. It is not hard to
find that the entropy is maximal when $\left\vert a\right\vert =\left\vert
b\right\vert =\frac{1}{\sqrt{2}}$. Thus the quantum channel described by Eq.
(\ref{es}) is optimal not only for the suppression of the phase noise, but
also for the transfer of the gravitational information.

\section{Discussion and Conclusion}

In this paper we propose that classical gravitational information can be
transferred through a quantum channel by a different method from quantum
teleportation, in which the quantum state \cite{bbw93} or the quantum field
\cite{ncc09} are transferred. In our proposal, the transferred information is
stored in the correlation between the internal and external degrees of freedom
of a single atom. It is interesting to note that similar situations such as
measurement of the gravitational constant G \cite{ffk07} or the general
relativistic proper time \cite{zcb11} could be implemented with our method.
Complete information transfer in our proposal is realised only by a partial
measurement (sometimes this is also called a local measurement, as opposed to
the non-local Bell measurement of quantum teleportation), i.e. we measure only
the external state of the second atom. This partial measurement only makes the
information shared between the two atoms, which is an obvious difference
between our proposal and standard quantum teleportation. However, the
information can be obtained by measuring either atom. More interesting, our
present proposal also applies to the \textquotedblleft cat
state\textquotedblright\ of many atoms, i.e. the gravitational information can
be read from any atom in the cat state as long as one atom of them goes
through the interferometer. It is stressed that the information can only be
read once even in such state of many atoms, but the potential transfer has be
realised among these atoms.

Due to the existence of a quantum channel, phase noise is suppressed, and thus
information transferred about the tested gravitational effects is improved in
contrast to the case without the quantum channel. This is an important reason
for choosing a quantum channel in our proposal, although a state without
quantum correlation \cite{hhs05,skb11}, $\left\vert \phi\right\rangle
=\frac{1}{2}\left(  \left\vert g_{1}\right\rangle \left\langle g_{1}%
\right\vert \otimes\left\vert e_{2}\right\rangle \left\langle e_{2}\right\vert
+\left\vert e_{1}\right\rangle \left\langle e_{1}\right\vert \otimes\left\vert
g_{2}\right\rangle \left\langle g_{2}\right\vert \right)  $, will also realise
the transfer of gravitational information when the second atom goes through
the interferometer. In particular, for the state $\left\vert \phi\right\rangle
$, an interesting effect is that when the second atom finishes the interaction
with the gravitational field, the new state $\left\vert \phi^{^{\prime}%
}\right\rangle $ will include the quantum correlation between them due to the
existence of cross terms such as $\left\vert g\right\rangle \left\langle
e\right\vert $. So one attractive open problem is of whether gravitational
interaction could generate quantum correlation for states such as $\left\vert
\phi\right\rangle $, in contrast with the analysis in Ref.
\cite{ld84,rp96,sla07} that suggests that gravity leads to the loss of
coherence of quantum states.

In summary, firstly we presented the \textquotedblleft specific
state\textquotedblright\ in the context of the background of the atom
interferometer first suggested by Kasevich and Chu, and analysed how
gravitational information is stored in such a state with correlation of the
internal and external degrees of freedom of the atom. Such a specific
information-carrying method makes the implementation of the transfer of
information by standard quantum teleportation unlikely. So in this paper we
have proposed a practical and novel scheme to transfer gravitational
information stored in a \textquotedblleft specific state\textquotedblright%
,\ and we have found that a complete transfer requires only a partial
measurement. Significantly, information transferred is best with respect to
phase noise for a quantum channel with maximal entanglement. Thus we have
presented a novel proposal for the transfer of classical gravitational
information using maximally entangled atomic pairs, which could be prepared in
current experiments. In particular, due to the improved and more precise
transfer of gravitational information, our proposal could also be regarded as
an improved readout mechanism for an atomic interferometer.

\section{Acknowledgement}

We are grateful to the anonymous referee for their critical comments and
helpful advice. This work is supported financially by National Basic Research
Program of China with Grant No. 2010CB832805 and National Natural Science
Foundation of China with Grant Nos. 11074283, 11104324.

\end{document}